# Bursting with Feedback: The Relationship between Feedback Model and Bursty Star Formation Histories in Dwarf Galaxies

Bianca Azartash-Namin[1,10], Anna Engelhardt[2,10], Ferah Munshi[2,3], B. W. Keller[4], Alyson M. Brooks[5,6],
Jordan Van Nest[1], Charlotte R. Christensen[7], Tom Quinn[8], and James Wadsley[9]
[1] University of Oklahoma, Department of Physics and Astronomy, Norman, OK 73019, USA; bianca.azartash.namin-1@ou.edu
[2] George Mason University, Department of Physics and Astronomy, Fairfax, VA 22030, USA
[3] Perimeter Institute, Waterloo, Ontario, Canada
[4] University of Memphis, Department of Physics and Materials Science, Memphis, TN 38152, USA
[5] Rutgers, The State University of New Jersey, Department of Physics and Astronomy, Piscataway, NJ 08854, USA
[6] Center for Computational Astrophysics, New York, NY 10010, USA
[7] Grinnell College, Department of Physics, Grinnell, IA 50112, USA
[8] University of Washington, Department of Astronomy, Seattle, WA 98195, USA
[9] McMaster University, Department of Physics and Astronomy, Hamilton, Ontario, Canada

## Abstract

Due to their inability to self-regulate, ultrafaint dwarfs are sensitive to prescriptions in subgrid physics models that converge and regulate at higher masses. We use high-resolution cosmological simulations to compare the effect of bursty star formation histories (SFHs) on dwarf galaxy structure for two different subgrid supernova (SN) feedback models, superbubble and blastwave, in dwarf galaxies with stellar masses from $5000 < M_*/M_\odot < 10^9$. We find that in the "MARVEL-ous Dwarfs" suite both feedback models produce cored galaxies and reproduce observed scaling relations for luminosity, mass, and size. Our sample accurately predicts the average stellar metallicity at higher masses, however low-mass dwarfs are metal poor relative to observed galaxies in the Local Group. We show that continuous bursty star formation and the resulting stellar feedback are able to create dark matter (DM) cores in the higher dwarf galaxy mass regime, while the majority of ultrafaint and classical dwarfs retain cuspy central DM density profiles. We find that the effective core formation peaks at $M_*/M_{\rm halo} \simeq 5 \times 10^{-3}$ for both feedback models. Both subgrid SN models yield bursty SFHs at higher masses; however, galaxies simulated with superbubble feedback reach maximum mean burstiness values at lower stellar mass fractions relative to blastwave feedback. As a result, core formation may be better predicted by stellar mass fraction than the burstiness of SFHs.

*Unified Astronomy Thesaurus concepts:* Dwarf galaxies (416); Dark matter density (354); Star formation (1569); N-body simulations (1083); Hydrodynamical simulations (767)

## 1. Introduction

In the Λ cold dark matter (ΛCDM) model of cosmological structure formation, dwarf galaxies are predicted to be the most abundant and yet the least luminous galaxies in the Universe. The ΛCDM framework reproduces the large-scale structure of the Universe well, however, at small scales dark matter (DM)-only simulations face tension with observations. One such tension is the cusp–core problem (Flores & Primack 1994; Moore 1994), which is described as the divergence from the Navarro–Frenk–White profile, and scales as $\rho(r) \propto r^\alpha$ with $\alpha \sim -1$, at small radii (Navarro et al. 1997). However, observations of dwarf galaxy rotation curves reveal that constant inner DM density "core" profiles ($\alpha \sim 0$) best describe the DM structure for subgalactic-scale galaxies (de Blok et al. 2008; Oh et al. 2011). DM-only simulations face many additional tensions with observations such as the missing satellite problem (Klypin et al. 1999; Moore et al. 1999) and the too big to fail problem (Boylan-Kolchin et al. 2011, 2012).

The inclusion of baryonic physics in ΛCDM simulations has been shown to resolve or greatly reduce tensions with observations such as the too big to fail and missing satellite problems (Brooks et al. 2013; Brooks & Zolotov 2014; Brook & Di Cintio 2015; Papastergis & Shankar 2016; Sawala et al. 2016; Wetzel & Hopkins 2016; Garrison-Kimmel et al. 2019; Munshi et al. 2021). Simulations that incorporate subgrid baryonic physics models also successfully predict cored profiles (Navarro et al. 1996; Gelato & Sommer-Larsen 1999; Read & Gilmore 2005; Oh et al. 2011; Pontzen & Governato 2012; Chan et al. 2015), resolving the cusp–core problem. Subgrid models of supernova (SN) feedback drive baryonic matter from the center of CDM halos, causing fluctuations in the gravitational potential and lowering the central density of DM (Governato et al. 2012; Zolotov et al. 2012; Di Cintio et al. 2014; Read et al. 2016). These fluctuations in the central gravitational potential indicate that both a supply of cold gas undergoing collapse and rapid expansion may be necessary to produce cored halos (Pontzen & Governato 2012).

Dwarf galaxies are susceptible to quenching due to internal feedback processes such as radiation, gas outflows, and winds (Governato et al. 2010). In particular, the process of gas removal from galaxies due to SN feedback is able to flatten cuspy density profiles to a core by lowering the central density of DM (Pontzen & Governato 2012). The stellar feedback mechanism driving galactic winds and outflows is therefore key to producing cores and necessary to model realistic dwarf galaxy star formation histories (SFHs) and DM responses.

The inclusion of subgrid models for baryonic processes creates more realistic galaxies (Stinson et al. 2007;

---

[10] Bianca Azartash-Namin and Anna Engelhardt are equal contributors to this work and designated as co-first authors.

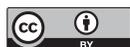

Mashchenko et al. 2008; Sawala et al. 2010, 2016; Governato et al. 2010, 2012; Pontzen & Governato 2012; Zolotov et al. 2012; González-Samaniego et al. 2014; Chan et al. 2015; Oñorbe et al. 2015; Wang et al. 2015; Wetzel & Hopkins 2016; Garrison-Kimmel et al. 2019; Hopkins et al. 2020; Engler et al. 2021; Font et al. 2021; Munshi et al. 2021). However, star formation is a complex process occurring on a large variety of timescales, making it difficult to model accurately. In a study of Milky Way–mass galaxies, feedback was found to make small-scale contributions to pressure support, which is crucial for regulating star formation and the vertical structure of the interstellar medium (ISM; Benincasa et al. 2016). Benincasa et al. (2016) found that realistic, regulated star formation requires that the scale height be resolved, allowing for subgrid models to effectively modulate star formation that is otherwise regulated in large-scale galaxies by feedback processes counterbalancing any changes to the star formation model.

There is evidence of self-regulation in galaxies with stellar masses as low as $M_* \gtrsim 10^{5-6}\ M_\odot$ in the "classical" dwarf galaxy regime (Stinson et al. 2007; Valcke et al. 2008), however, self-regulation is restricted to the regime of strong feedback (Semenov 2019). Strong feedback refers to high star formation efficiency and efficient feedback, meaning that the rate of gas removal by feedback is comparable to the rate of star formation. Ultrafaint dwarfs (UFDs) are believed to lie below the mass regime of strong feedback meaning that they are unable to self-regulate and may be highly sensitive to prescriptions of star formation and feedback. Munshi et al. (2019) found that adjusting the star formation recipe in galaxies with $M_{vir} < 10^9\ M_\odot$ produced drastically different results, indicating that galaxies that are too small to self-regulate are extremely sensitive to the choice of subgrid model. Agertz et al. (2020) show that the stellar mass–metallicity relation is highly sensitive to the strength of SN feedback at the UFD scale.

The blastwave feedback model (McKee & Ostriker 1977; Stinson et al. 2006) incorporates individual SN blastwaves as stellar feedback to appropriately regulate star formation for galactic smoothed particle hydrodynamic (SPH) simulations. Blastwave feedback solves the numerical overcooling problem by limiting the loss of thermal energy from SNe through radiative cooling in the surrounding gas particles by temporarily turning cooling off through the snowplow phase (see also Brook et al. 2004). Simulations implementing blastwave feedback are able to reproduce observed star formation rates and stellar masses (Stinson et al. 2006; Guedes et al. 2011; Munshi et al. 2013, 2021), though Keller et al. (2015) find that blastwave feedback alone does not sufficiently regulate star formation in $L^*$ galaxies.

Blastwave models are resolution dependent and do not account for physical processes that occur in nature such as clustered star formation (Nath & Shchekinov 2013; Sharma et al. 2014). As star formation is clustered, feedback from the individual winds and SN energy merge, thermalize, and form superbubbles. The superbubble feedback model introduced in Keller et al. (2014) uses thermal conduction to produce the expected interior densities of superbubbles by accounting for temperature gradients, evaporation processes, and implementing a separate hot and cold phase treatment of gas to avoid overcooling. Keller et al. (2014, 2015) and Keller (2022) found that the superbubble feedback model shows stronger star formation regulation by reducing star formation rates by a factor of 2 relative to the blastwave model and is more effective at driving mass-loaded outflows by roughly an order of magnitude. This feedback model is potentially a solution for modeling galaxies within the ΛCDM framework without the use of unphysical cooling shutoffs.

Keller et al. (2014, 2015) and Keller (2022) find that simulations run with a blastwave feedback model produce halos with overmassive stellar populations while the superbubble feedback model produces more massive outflows that efficiently expel low angular momentum gas from the centers of Milky Way–mass galaxies. Effective expulsion of low angular momentum gas particles at high redshift limits star formation at the centers of galaxies, subsequently avoiding the creation of overly massive stellar bulges (Governato et al. 2010; Brook et al. 2011; Christensen et al. 2014). As a result, superbubble galaxies are more disk dominated than blastwave galaxies (Keller et al. 2015) and have less centrally dense stellar populations (Keller et al. 2015; Keller 2022).

In this paper, we will investigate the impact of stellar feedback models in dwarf galaxy simulations by addressing the differences in subgrid physics for the blastwave and superbubble feedback models and how repeated bursts of star formation impact the formation and sustainability of DM cores within these models. Mina et al. (2021) found within a small sample of dwarf galaxy masses that the superbubble model implements the physics behind SN feedback processes more accurately than the delayed cooling shutoff for the blastwave model. This subsequently reproduces reasonably realistic dwarf galaxies with comparable stellar and gas properties that are observed within the Local Volume. We expand upon this treatment of stellar feedback implementation by directly calculating core slopes and burstiness values of the SN rate using a parameter defined in Section 3.3 to represent bursty star formation. We calculate burstiness and core slopes for both feedback models and sample a full range of dwarf galaxy masses using zoomed-in simulations that contain dozens of dwarfs ranging from Large Magellanic Cloud (LMC) mass down to the UFD mass regime. Additionally, we adopt a different prescription for star formation which is physically motivated by the dependence on local nonequilibrium abundance of molecular hydrogen as opposed to relying on the simple density and temperature cutoff used in Mina et al. (2021).

In Section 2 we describe our simulations, providing an explanation of key differences between the two feedback models we compare in this analysis: blastwave and superbubble. In Section 3.1 we study the observable properties of galaxies simulated using the superbubble and blastwave feedback mechanisms. We discuss the implications of our results in Section 4 and summarize our work in Section 5.

## 2. Simulations

For the comparison of bursty feedback and DM core formation, we use a sample of dwarf galaxies from the "MARVEL-ous Dwarfs" (MARVEL) zoom-in simulations, in particular, one of the four volumes we call Storm. These MARVEL simulations generate a large sample of simulated dwarf galaxies at a high resolution (60 pc force resolution), with gas, initial star, and DM masses of $1410\ M_\odot$, $420\ M_\odot$, and $6650\ M_\odot$ respectively, which allow for resolving galaxy masses as low as $M_{star} \sim 6000\ M_\odot$ (Munshi et al. 2021). We analyze galaxies from two separate runs of the Storm simulation: one that implements blastwave feedback and a second that implements superbubble feedback. Our sample consists of a total of 19 dwarfs in the superbubble run and 21 in the blastwave run.

The simulations were run from $z = 149$ to $z = 0$ within a Wilkinson Microwave Anisotropy Probe cosmology (Spergel et al. 2007). The regions are located approximately 1.5–7 Mpc away from a Milky Way–mass galaxy and can be thought of as a representation of the Local Volume (Munshi et al. 2021). The simulation is run with the $N$-body + SPH code CHANGA (Menon et al. 2015) to scale up to thousands of cores via a tree-based gravity solver and the CHARM++ runtime system (Kale & Krishnan 1993).

The Storm simulations, as with all the MARVEL simulations, reproduce a Schmidt Law (Schmidt 1959; Kennicutt 1998) with probabilistic star formation that depends on the local $H_2$ abundance, gas density, and gas temperature as described in Christensen et al. (2012). Star particles represent stellar populations with a Kroupa (2001) initial mass function (IMF) and initial mass that is 30% that of the original gas particle. Star formation efficiency (the fraction of a gas particle converted into stars at a local dynamical time) is calculated using the following equation

$$c^* = 0.1 f_{H_2} = 0.1 \frac{x_{H_2}}{x_{H_2} + x_{HI}}, \quad (1)$$

where $x_{H_2}$ and $x_{HI}$ are the mass fractions of molecular hydrogen and atomic hydrogen, respectively. Star formation is additionally constrained to gas particles with temperatures less than $10^3$ K and densities $\rho > 0.1$ amu cm$^{-3}$. The simulations include dust and self-shielding of $H_2$ as well as dust shielding of H I, where the column length is calculated using a particle smoothing length of 6 pc. Observational evidence supports a higher star formation efficiency in gas surface densities where molecular hydrogen dominates the gas profile (Christensen et al. 2012). As a result, star formation is constrained to areas with ample $H_2$. Additionally, gas cooling of $H_2$ is integral to the production of a cold ISM where efficient star formation takes place (Christensen et al. 2012). Supermassive black hole formation, growth, mergers, feedback, and dynamics are also based on local gas conditions (Bellovary et al. 2011; Tremmel et al. 2015, 2017; Bellovary et al. 2019).

The simulations implement metal cooling (MC) and diffusion (Shen et al. 2010) with a time-dependent uniform UV background (Haardt & Madau 2012). The UV background decreases the MC rate for temperatures above $10^4$ K, but below this threshold the cooling rates are increased due to a larger number of free electrons. Shen et al. (2010) found that galactic winds most efficiently enrich the intergalactic medium (IGM) for intermediate-mass galaxies between $10^{10}$ and $10^{11} M_\odot$, and lower-mass galaxies remain metal poor because gas is prevented from efficiently accreting. The process of metal diffusion decreases the metal content in the IGM by mixing winds before they escape, increasing the quantity of low-metallicity gas contained within these winds (Shen et al. 2010).

The Storm volume is run twice, each with a distinct feedback model. The first is the blastwave SN feedback model (Stinson et al. 2006) in which the nearby gas particles are directly injected with mass, metals, and a fixed fraction of the total energy from each SN in the form of thermal energy at $1.5 \times 10^{51}$ erg per SN event. The number of SNe is calculated from the IMF of each star particle and it is assumed that only stars with masses ranging from 8 to 40 $M_\odot$ undergo Type II SNe. Cooling in particles within a blast radius is temporarily turned off through the snowplow phase to prevent the numerical overcooling of energy deposited in the local ISM (Stinson et al. 2006). The number of particles with cooling disabled depends on an analytic treatment of blastwaves (McKee & Ostriker 1977). Particles outside of the blast radius have enhanced mass and metallicity, but experience immediate radiative thermal cooling.

The second model is the superbubble feedback model (Keller et al. 2014), which introduces thermal conduction, subgrid evaporation, and subgrid multiphase treatment to model the early stages of superbubbles in the case of low-mass resolution within the simulation. Evaporation from thermal conduction accurately regulates the quantity of hot gas without the addition of an unphysical cooling shutoff and the model is insensitive to resolution. Thermal conduction maintains uniform temperatures in hot bubbles while a two-phase treatment of gas particles prevents overcooling and is used for cases where the simulation mass resolution is insufficient to model the early stages of the superbubble. We follow the treatment of energy per SN event outlined by Keller et al. (2015) and reproduced by Mina et al. (2021) to follow the heating time associated with Type II SNe from OB stars with $10^{51}$ erg per SN event. As a result, the superbubble model initially injects less thermal energy into the surrounding gas particles than the blastwave model.

With this star formation prescription, the MARVEL dwarf galaxy simulations are able to produce cored DM density profiles. In the following sections, the selected galaxies from the MARVEL Storm blastwave and superbubble feedback runs have a minimum of 14 star particles and a minimum extended SFH of 100 Myr. Satellite halos are excluded from our sample to limit the impact of external mechanisms such as tidal stirring on star formation and gas removal. The halos are identified with Amiga's Halo Finder (AHF; Gill et al. 2004; Knollmann & Knebe 2009).

The virial radius, $R_{\rm vir}$, is defined as the radius at which the average halo density is 200 times the critical density of the universe at a given redshift, 200 $\rho_{\rm crit}(z)$. Halo properties are calculated within the virial radius using the AHF catalog and particles within the virial radius are considered bound to the halo unless the particle velocity exceeds the escape velocity. The $V$-band luminosity is calculated by finding the visual magnitude of every star passed in the halo. PYNBODY includes a grid of simple stellar population luminosities from Girardi et al. (2010; see also Marigo et al. (2008)) for many bandpasses and various stellar ages and metallicities. This function linearly interpolates to the desired value and returns the value as a magnitude. The magnitudes of each individual star particle in the halo are converted into luminosities and added together. The half-light radius is computed using a function which calculates the entire luminosity of star particles residing within the halo, finds half that, sorts stars by distance from the halo center, and finds out inside which radius the half luminosity is reached. We do not include dust corrections, however, because dust content broadly scales with stellar mass (Gelli et al. 2021), thus it is unexpected for the intrinsic spectrum to be significantly attenuated in dwarf galaxies. The PYNBODY analysis code (Pontzen et al. 2013) was used for further analysis.

Halos are considered to be matched between the superbubble and blastwave runs of the simulations by determining the percentage of shared DM particles. Halos are matched if more than 75% of the DM particles in a given superbubble halo share the same particle ID as the DM particles in the corresponding halo from the blastwave run. Not all halos have a matched counterpart.

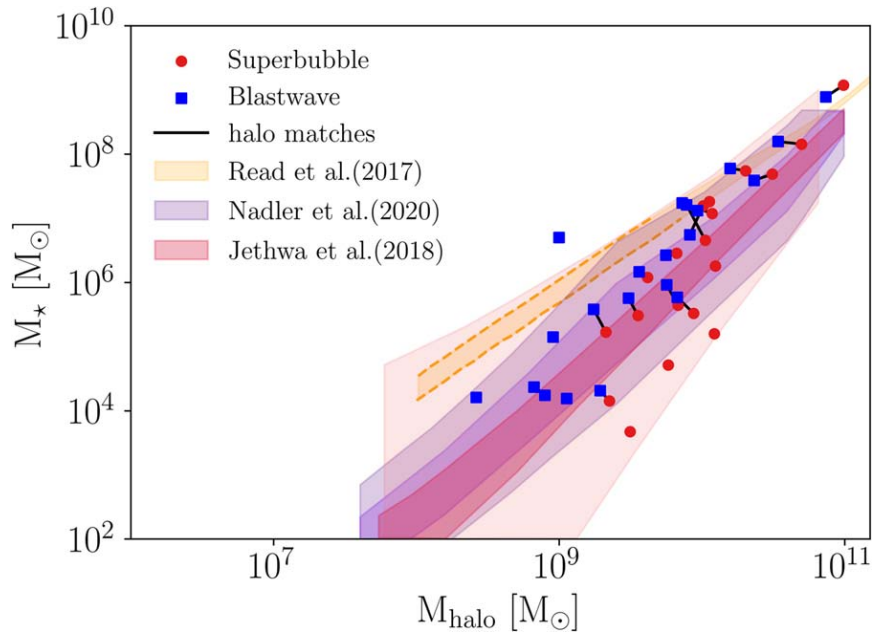

**Figure 1.** Stellar mass vs. halo mass for the Storm blastwave and superbubble feedback models. Red circles represent galaxies simulated using superbubble feedback while blue squares represent galaxies simulated using blastwave feedback. Black lines connect the corresponding halos across simulations, while unconnected halos do not have a matched counterpart. The orange region is the SDSS field stellar mass function from Read et al. (2017) with dashed orange lines denoting where it relies on a power-law extrapolation. The red (Jethwa et al. 2018) and purple (Nadler et al. 2020) contours correspond to $P(M_*|M_{\rm vir})$ constraints with darker colors representing the 68% confidence interval and lighter colors representing the 95% confidence interval.

## 3. Results

### 3.1. Observational Comparison

The following section investigates ways to observationally distinguish between the superbubble and blastwave feedback models. Future surveys performed by the Vera Rubin Observatory (VRO) and Nancy Grace Roman Space Telescope are projected to discover dozens of UFD galaxies in the Local Group, and our analysis could help contextualize these new observations. We compare simulation data with the half-light radius, $V$-band luminosity, stellar metallicity, and gas-phase metallicity of galaxies in the local Universe. These galaxy properties were chosen because they remain some of the most accurate measurements in difficult-to-detect low-luminosity objects and can provide valuable insight into a galaxy's SFH. Galaxy properties from our sample are also compared with results from a variety of other simulations, including the Seven Dwarfs simulations (Mina et al. 2021; Shen et al. 2014). Our study of the observational distinctions between the blastwave and superbubble SN feedback models is an extension of the analysis done by Mina et al. (2021). However, it is important to note that the Seven dwarfs and MARVEL dwarf simulations use different star formation prescriptions that result in different modeling of the ISM. Star formation in the Seven Dwarfs simulations is constrained to temperature and density thresholds of $T < 10^4$ K and $n_H > 100$ atoms cm$^{-3}$, respectively, while star formation in the MARVEL dwarfs is dependent on the local $H_2$ abundance. In an apples-to-apples comparison of the MC star formation model used by Mina et al. (2021) and the $H_2$ star formation model, Munshi et al. (2019) find that the range of densities in the $H_2$ run is higher than the MC run. The range of densities in the MC run is only 100−1000 atoms cm$^{-3}$ while the range in the $H_2$ run is 100−5.2 × 10$^5$ atoms cm$^{-3}$.

The stellar mass–halo mass (SMHM) relation for superbubble halos (red circles) and blastwave halos (blue squares) is shown in Figure 1. Both the superbubble and blastwave feedback models roughly exhibit a power-law relationship between stellar and halo mass that is tighter at higher masses, experiencing some degree of scatter at the low-luminosity end. Storm halos are well fit by the Sloan Digital Sky Survey (SDSS) field stellar mass function (Read et al. 2017) above $M_* \sim 10^7 \, M_\odot$ but generally lie below the power-law extrapolation of the stellar mass function at lower masses. Nearly all the galaxies in our sample are consistent with Jethwa et al. (2018) and Nadler et al. (2020). The power-law SMHM relation published by Jethwa et al. (2018) comes from a fiducial SMHM model that uses an informative prior on the stellar mass of a $10^{11} \, M_\odot$ halo based on abundance matching studies at higher mass scales. The SMHM relation published by Nadler et al. (2020) is inferred from fitting data from the Dark Energy Survey and the Panoramic Survey Telescope and Rapid Response System.

Governato et al. (2010) and Brook et al. (2011) find that the removal of low angular momentum gas at high redshift prevents the formation of bulges and contributes to the production of disk-dominated galaxies detected by observations. Keller et al. (2015) find that superbubble-driven outflows remove gas from the centers of $L^*$ galaxies more efficiently than blastwave feedback, resulting in galaxies that are bluer and more disk dominated than blastwave galaxies. The difference in star formation efficiency of the two models is demonstrated in Figure 1, where simulated blastwave halos with halo masses below $10^{10} \, M_\odot$ have elevated stellar masses.

The left-hand side of Figure 2 compares the average galaxy stellar metallicity at $z = 0$, measured by [Fe/H], plotted against luminosity in the visual $V$ band for our sample of simulated galaxies and observed galaxies in the Local Group (McConnachie 2012), as well as galaxies from the Seven Dwarfs simulations (Mina et al. 2021; Shen et al. 2014). As before, galaxies from the Storm simulation runs are colored dark blue for blastwave and red for superbubble. Galaxies in the superbubble and blastwave runs of the Seven Dwarfs simulations are colored pink and light blue, respectively. Despite using the same feedback models, the Seven Dwarfs simulations adopt a different star formation prescription than we use here. Observed galaxies are colored gray and include uncertainty measurements. Purple and green triangles represent galaxies from other simulations (Macciò et al. 2017; Revaz & Jablonka 2018; Wheeler et al. 2019; Agertz et al. 2020; Sanati et al. 2023) for comparison. The unfilled downwards pointing green triangles are an extension of the work from Revaz & Jablonka (2018), which was later published by Sanati et al. (2023). The orange pentagons represent galaxies from the DC Justice League (DCJL) simulations (Applebaum et al. 2021). The DCJL simulations provide a large sample of satellite dwarf galaxies simulated in an environment with a Milky Way–mass host halo. They use the same star formation prescription as the Storm runs in this work, but adopt the blastwave feedback model. Following the work of Hopkins et al. (2018), Agertz et al. (2020), and Applebaum et al. (2021) Population III stars are assumed to preenrich gas to a minimum metallicity and we impose a primordial metallicity floor of $Z > 10^{-5}$ for individual star particles. The implementation of this metallicity floor enables an apples-to-apples comparison of average stellar metallicities between the DCJL and MAVELous dwarf simulations.

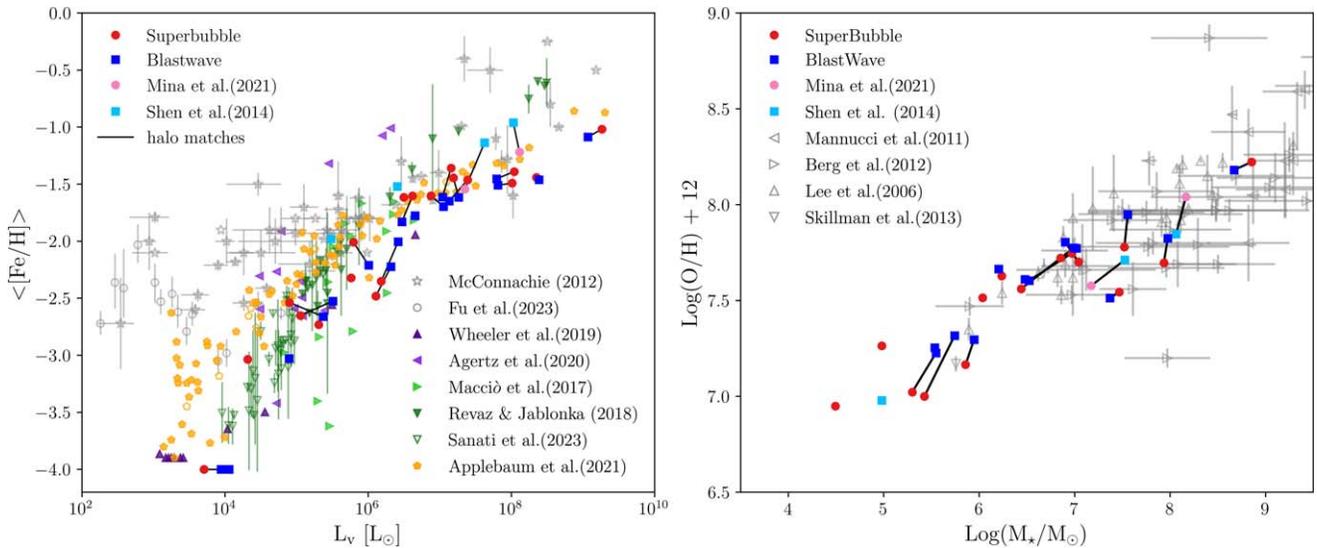

**Figure 2.** Left: relation between stellar metallicity and *V*-band luminosity for simulated and observed galaxies. As in Figure 1, blue squares and red circles represent Storm simulations evolved using the blastwave and superbubble feedback models, respectively. Light blue squares (blastwave) and pink circles (superbubble) represent halos in the Seven Dwarfs simulations (Mina et al. 2021; Shen et al. 2014). Black lines link the corresponding halos across simulations. Metallicity measurements of observed galaxies in the Local Group from McConnachie (2012) and Fu et al. (2023) are represented by empty gray stars and circles, respectively. Purple and green triangles represent halos from other simulations (Macciò et al. 2017; Revaz & Jablonka 2018; Wheeler et al. 2019; Agertz et al. 2020; Sanati et al. 2023) and the orange pentagons are from a sample of satellite/backsplash (filled in points) and isolated (empty points) halos in the DCJL simulations (Applebaum et al. 2021). Right: relation between gas-phase metallicity and stellar mass for simulated galaxies at $z = 0$. Storm and Seven Dwarfs galaxies are represented using the same markers from the left-hand side. Empty gray triangles represent galaxies in the Local Group (Mannucci et al. 2011; Berg et al. 2012; Lee et al. 2006; Skillman et al. 2013).

Average stellar metallicity provides insight into the SFH of an entire galaxy. Low-mass galaxies are systems with smaller stellar populations resulting in, on average, decreased luminosity and metallicity measurements relative to high-mass galaxies. The simulated galaxies with the lowest luminosities have systematically lower average stellar metallicities than the satellites in the Local Group, and this result is consistent across simulations. However, Applebaum et al. (2021) noted that, for the DCJL simulations, the discrepancy disappears when using total metallicity. This may indicate a problem with the Fe yields used in the simulations.

Mina et al. (2021) find that in the Seven Dwarfs volume simulations the superbubble feedback model produces galaxies with decreased metallicity relative to galaxies simulated using blastwave feedback (note that only two galaxies form in the superbubble run of the Seven Dwarfs, relative to four galaxies in the original blastwave simulation). We do not detect any clear visible distinctions between the blastwave and superbubble feedback models in the Storm galaxies. Additionally, we find that Storm galaxies have lower metallicities than those in the Seven Dwarfs blastwave runs. Since these simulations have similar feedback models but different star formation prescriptions, this difference may be a result of the differing star formation models.

The right-hand side of Figure 2 illustrates the relationship between gas-phase metallicity and stellar mass. The colors and shapes of each simulated data point are the same as on the left. All observations share the same gray color but are denoted by triangles with different orientations instead of stars. All of the simulated galaxies shown match the observations well. 10 halos from our sample (seven blastwave and three superbubble) lack cold gas and, as a result, have no corresponding gas-phase metallicity calculations. There are no obvious distinctions between the feedback models in the Storm or Seven Dwarfs simulations, and the Storm galaxies show close agreement with the Seven Dwarfs galaxies despite the different ISM and star formation prescriptions.

Figure 3 plots stellar half-light radius against stellar mass and galaxies are denoted in the same manner as in Figure 2. The half-light radius is calculated by measuring the total luminosity of a given galaxy, then finding the radial distance within which half of the total luminosity is contained. The simulation data are compared to galaxies in a range of environments from the Local Group (McConnachie 2012). In a comparison by Mina et al. (2021), superbubble galaxies were found to have an increased half-light radius relative to blastwave galaxies. However, in the Storm simulation runs there is no apparent difference in the half-light radius between the feedback models. Again, the feedback models are the same in the Storm and Seven Dwarfs simulations and the main difference is in the star formation prescription. It may be that the lack of a clear impact on size in the Storm runs is due to the adopted star formation prescription rather than the feedback model. The sizes of the Storm galaxies are in good agreement with the observations of the Local Group. In contrast, the simulated galaxies from Wheeler et al. (2019) have increased half-light radii relative to the observations.

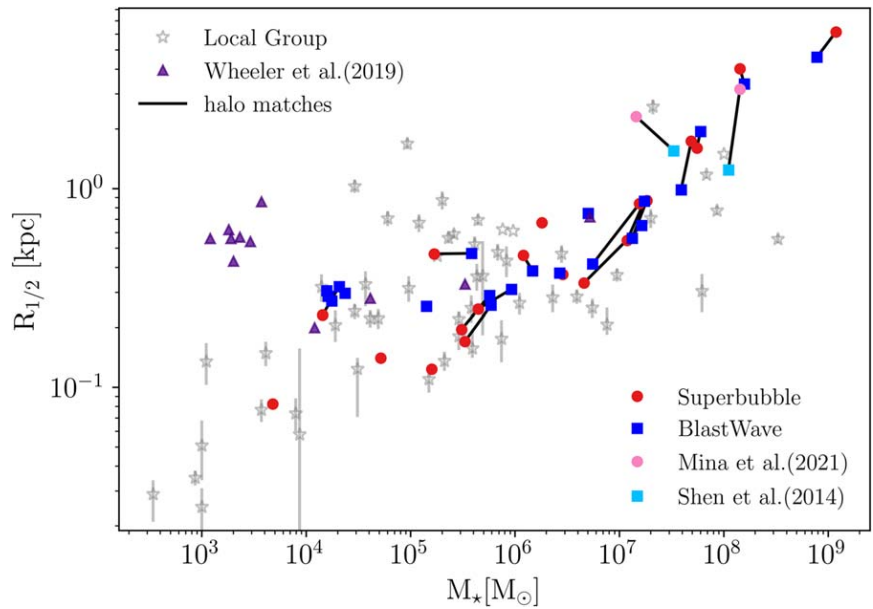

**Figure 3.** Relation between half-light radius and stellar mass for simulated galaxies at redshift $z = 0$. As before, dark blue squares and red circles represent Storm simulations evolved using the blastwave and superbubble feedback models, respectively. Light blue squares (blastwave) and pink circles (superbubble) represent halos in the Seven Dwarfs simulations (Shen et al. 2014; Mina et al. 2021). Black lines link the corresponding halos across simulations. Empty gray stars with error bars represent observed galaxies in the Local Group (McConnachie 2012). Purple upward pointing triangles represent resolved central galaxies at high resolution from Wheeler et al. (2019).

### 3.2. Dark Matter Cores

Under certain conditions, fluctuations in the gravitational potential can result in an DM density inner profile that is cuspy, flattening to a more cored profile. Pontzen & Governato (2012) find that rapidly expanding gas as a result of "bursty" star formation allows models to reveal central potential fluctuations capable of generating DM cores. Sales et al. (2022) discuss how DM core formations in cosmological simulations are commonly revealed when resolving star formation with high gas density thresholds that also result in bursty star formation.

Dutton et al. (2019) show that star formation becomes burstier and more clumped for higher star formation density thresholds. However, the burstiness of star formation alone is not sufficient for core formation. Instead, core formation occurs when high gas densities are resolved and star formation is restricted to occur at high densities. Jahn et al. (2023) find cores are capable of forming when the star formation density thresholds are set low, but affirm that cores only form when a multiphase ISM that resolves sufficiently high density contrast is resolved. All of the blastwave and superbubble simulations in this work adopt star formation prescriptions that have been shown to create DM cores (Governato et al. 2010, 2012; Shen et al. 2014; Mina et al. 2021). The range of temperatures for gas particles in the blastwave run of the simulation is $10 < T < 2.3 \times 10^6$ K and $10 < T < 6.8 \times 10^4$ K for the superbubble run. Gas particle densities range from $10^{-14}$ to 435 amu cm$^{-3}$ at redshift $z = 0$. However, the ability to form a DM core also depends on the stellar mass of the galaxy, as the stellar mass sets how much SN feedback is available to induce fluctuations in the gravitational potential. A number of studies have found that inner DM density slopes depend strongly on the mass ratio of stellar mass to total halo mass, $M_*/M_{halo}$. In general, most simulations find a characteristic mass ratio for peak core formation at $M_*/M_{halo} \simeq 5 \times 10^{-3}$ (Di Cintio et al. 2014; Chan et al. 2015; Tollet et al. 2016; Lazar et al. 2020), where values above and below this fall back into the the "cuspy" density profiles that are prevalent with DM-only simulations (see Figure 4).

To address the slowly rising rotation curves of dwarf galaxies that are best characterized by cored profiles, Lazar et al. (2020) introduced a three-parameter core–Einasto profile that includes a radius–core parameter ($r_c$) that defines where a density profile begins to flatten into a constant-density core. We adopt the core–Einasto profile to analyze the inner DM density slopes ($\alpha$) of the Storm halos. For the halos presented in this paper, $\alpha = 0$ for cored profiles, and slopes are calculated from the best-fit core–Einasto model at 1%–2% $R_{vir}$.

Figure 4 plots the DM density slopes in the Storm simulations against $M_*/M_{halo}$. Black data points represent halos from the FIRE-2 simulations (circles) and DM-only simulations (crosses; Lazar et al. 2020). While all of the compiled simulations find that DM core creation is maximized for $M_*/M_{halo} \sim 5 \times 10^{-3}$, above and below this $M_*/M_{halo}$ value the core slopes become more negative (cuspy) at different rates for different models. The superbubble and blastwave feedback models are indistinguishable from each other. They have flatter core-slope values relative to the halos studied in Lazar et al. (2020), and the slopes steepen less quickly with declining $M_*/M_{\rm halo}$. The data from the Storm simulations best match the results from the Di Cintio et al. (2014) profile, though the trend also follows a similar shape to Tollet et al. (2016) but with flatter DM density slopes. In contrast the central cores of DM-only halos are always cuspy, indicating that the incorporation of baryonic physics significantly mitigates tensions with the cusp–core problem.

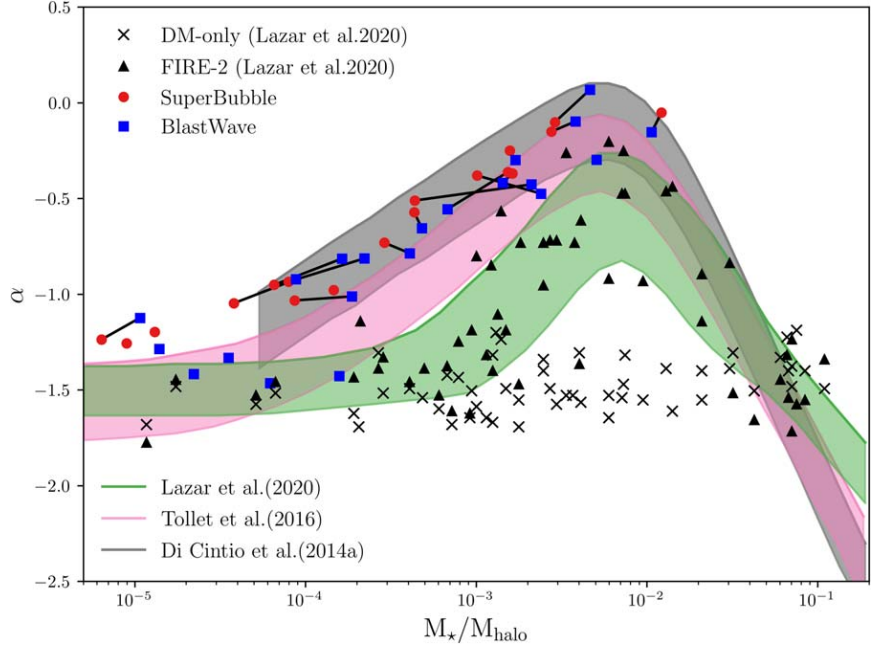

**Figure 4.** DM density profile slopes vs. $M_*/M_{\rm halo}$ at $z = 0$. Red circles and blue squares represent core slopes calculated for halos that use superbubble feedback and blastwave feedback models, respectively, with black lines linking the matched halos. These data are plotted with results from Di Cintio et al. (2014), Lazar et al. (2020), and Tollet et al. (2016) for comparison. Black points are halos from the FIRE simulations with triangles incorporating baryonic physics, and x symbols represent halos that come from DM-only simulations (Lazar et al. 2020). All slopes in this plot are measured at 1%–2% $R_{\rm vir}$.

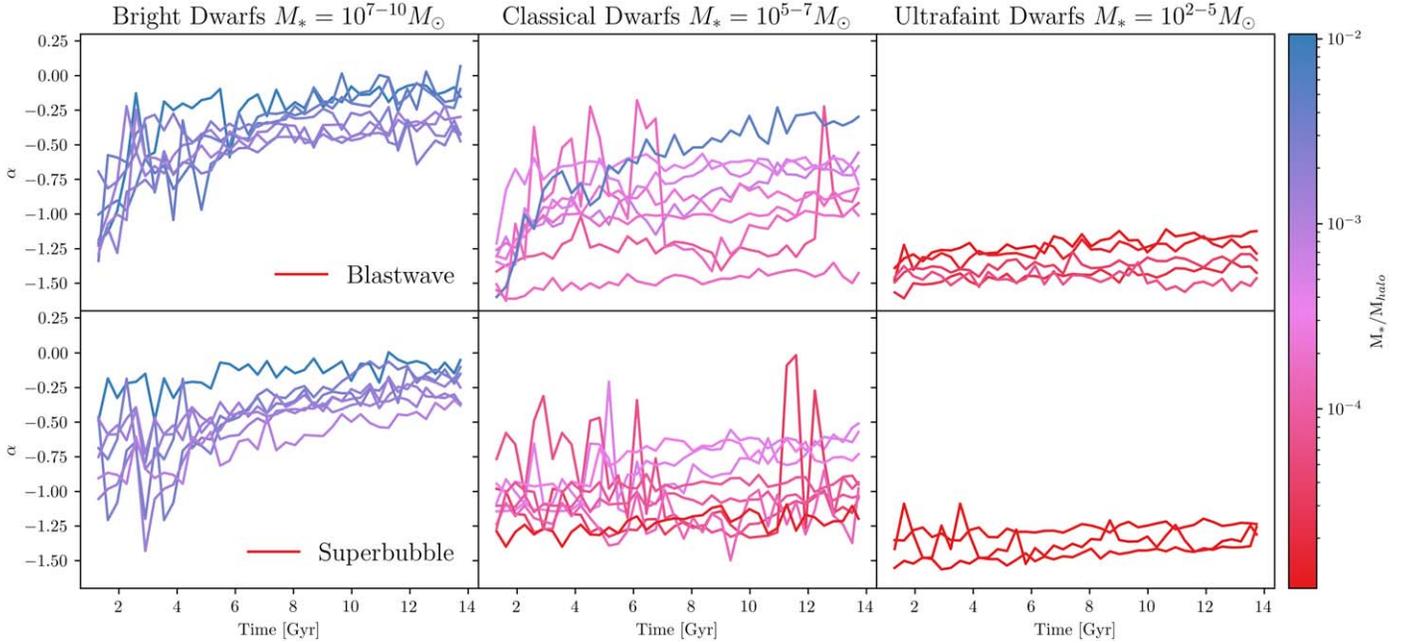

**Figure 5.** Core–Einasto fit slopes plotted through time [units of Gyr]. The top row corresponds to halos run using blastwave feedback while the bottom row shows halos run using superbubble feedback. Plots are colored by $M_*/M_{\rm halo}$ and columns separate halos by stellar mass. The leftmost column contains the most massive dwarfs ($10^7$–$10^{10}\ M_*[M_\odot]$), the middle column contains dwarfs in the stellar mass range $10^5$–$10^7\ M_*[M_\odot]$, and the least massive dwarfs ($10^2$–$10^5\ M_*[M_\odot]$) are shown in the rightmost column. These stellar mass ranges are classified as bright, classical, and ultrafaint, respectively, in accordance with Lazar et al. (2020).

Figure 5 demonstrates how higher-mass dwarfs, mostly limited to the bright dwarf regime, form cores over time while the majority of classical and UFDs do not experience enough sustained star formation to form cores. Ultrafaints are shown to be extremely cuspy throughout their lifetimes while galaxies in the classical mass regime usually only form temporary cores before reverting to cuspy central density profiles. Jackson et al. (2023) find that cores are created in stages through repeated bursts of star formation in galaxies where the central mass of gas is comparable to that of the DM. The formation of cores in the Storm simulations is similarly restricted to halos which are more gas rich. We found that at least 1% of the total halo mass was composed of gas for Storm halos with core-slope values above −0.5 in both feedback runs.

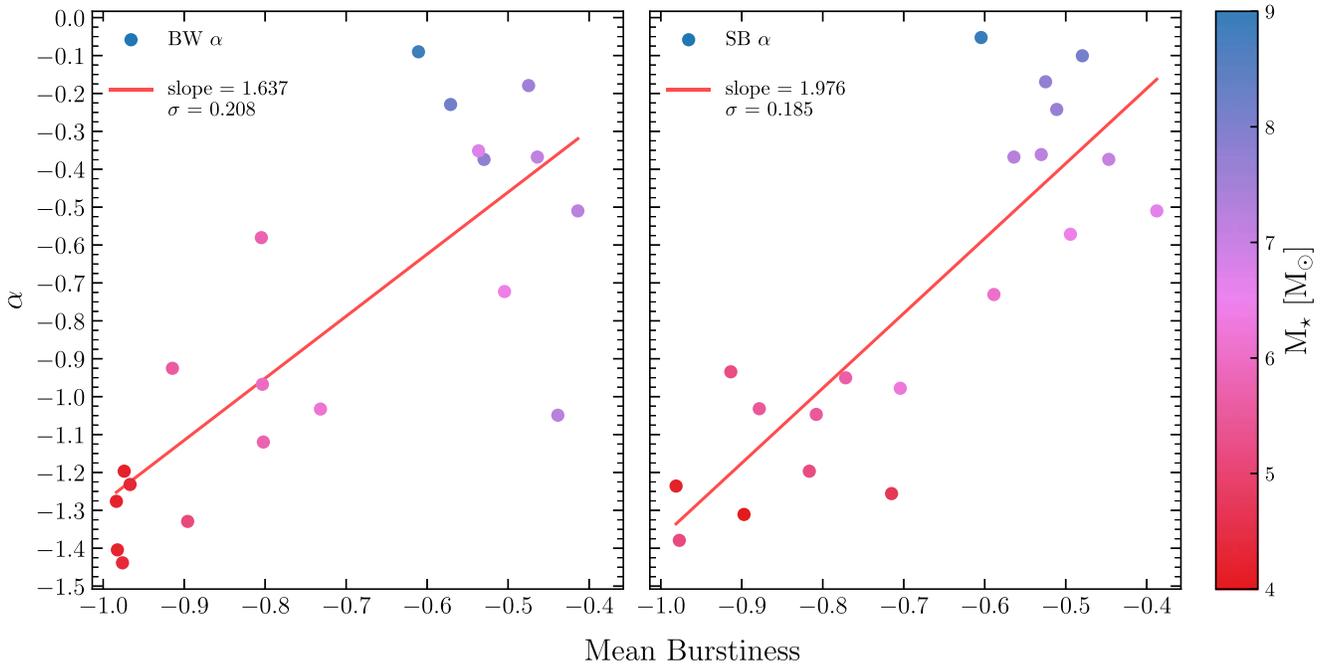

**Figure 6.** $\alpha[1\%–2\%]r_{\rm vir}$ core–Einasto fit core slopes as a function of burstiness for the Storm dwarf galaxies. Plots are colored by $M_\star\,[M_\odot]$ and columns separate halos by feedback model. The standard deviation is calculated to quantify which feedback model has less scatter, resulting in lower $\sigma = 0.19$ for the superbubble model.

### 3.3. Quantifying the Burstiness of Star Formation Histories

Dwarf galaxy simulations that include baryonic physics predict that feedback-induced DM cores result from bursty star formation driving rapid gas outflows and core formation is sustained with multiple starburst cycles. (Read & Gilmore 2005; Pontzen & Governato 2012). The following section investigates the relationship between burstiness and DM core formation. We investigate the Storm halos over a Hubble time by using the burstiness parameter $B = \frac{\sigma/\mu - 1}{\sigma/\mu + 1}$, where $\sigma$ is the standard deviation of the SN rate and $\mu$ is the mean SN rate (see also Applebaum et al. 2020). Thus burstiness ranges from $-1$ to $1$ where a uniform distribution, or the SN rate $= 0$, has a burstiness $B = -1$ and an exponential distribution has a burstiness $B = 0$. Burstiness will approach 1 as $\sigma/\mu \to \infty$. The SN rate is calculated in 1 Myr increments for the age of the universe for all halos within the simulation. The rate is calculated for the first 50 Myr of star formation to coincide with the approximate cooling time for SN feedback in our simulations. The burstiness of the SN rate is averaged over this windowed timescale of 50 Myr from the first star-forming particle to the current age of the universe. This timescale is selected to be on order of the scale of the feedback timescale, or put another way, the typical age of an O or B star. For halos that have truncated SFHs (in particular, UFDs) we find that $B$ is robust to the averaging timescale; for halos with continuous SFHs, we find that $B$ is in fact sensitive to the averaging timescale. However, for values of the averaging timescale that vary within 10–500 Myr, $B$ changes by only a factor of 2. Timescales the length of 1 Gyr and longer result in all halos having nonbursty $B$ values.

The burstiness of the SN rates for each halo is then averaged over a windowed timescale corresponding to the cooling shutoff time for feedback in our simulations to directly compare burstiness and core slopes relative to the stellar mass fraction, $M_\ast/M_{\rm halo}$. The burstiness of the Storm halos for both the blastwave and superbubble feedback models is plotted with the corresponding stellar mass and core slope to determine the significance of sustained bursty star formation and DM core slopes in Figure 6.

In contrast with the burstiness calculated over a Hubble time, we define "active burstiness" within the timescale of active star formation for a given halo. Burstiness calculations are confined to time bins between the formation of the first star particle and the last star formation. The burstiness for each halo during active star formation is directly compared to the burstiness for the cosmological timescale in Figure 7.

Instantaneous burstiness is calculated as the burstiness within a selected $\Delta t$ between the last 100 Myr of our simulations. The last two burstiness measurements of the simulation are are separated by 50 Myr and averaged as a proxy for an instantaneous burstiness measurement at $z = 0$. We find no direct correlation between instantaneous burstiness and core slopes for the entire range of halo masses, although it is worth noting that there is a distinct mass cutoff for the ultrafaint mass regime. Instantaneous burstiness appears to be reflective of the SFH in the few megayears prior to the last time step, indicating that the UFD galaxies in the Storm simulation become quenched early on and do not have sustained bursty star formation throughout the chosen timescale.

A relationship between burstiness and $\alpha$ DM cores as a function of halo mass becomes evident in the Storm DM halos. The burstiness values increase with $M_\ast/M_{\rm halo}$ and directly

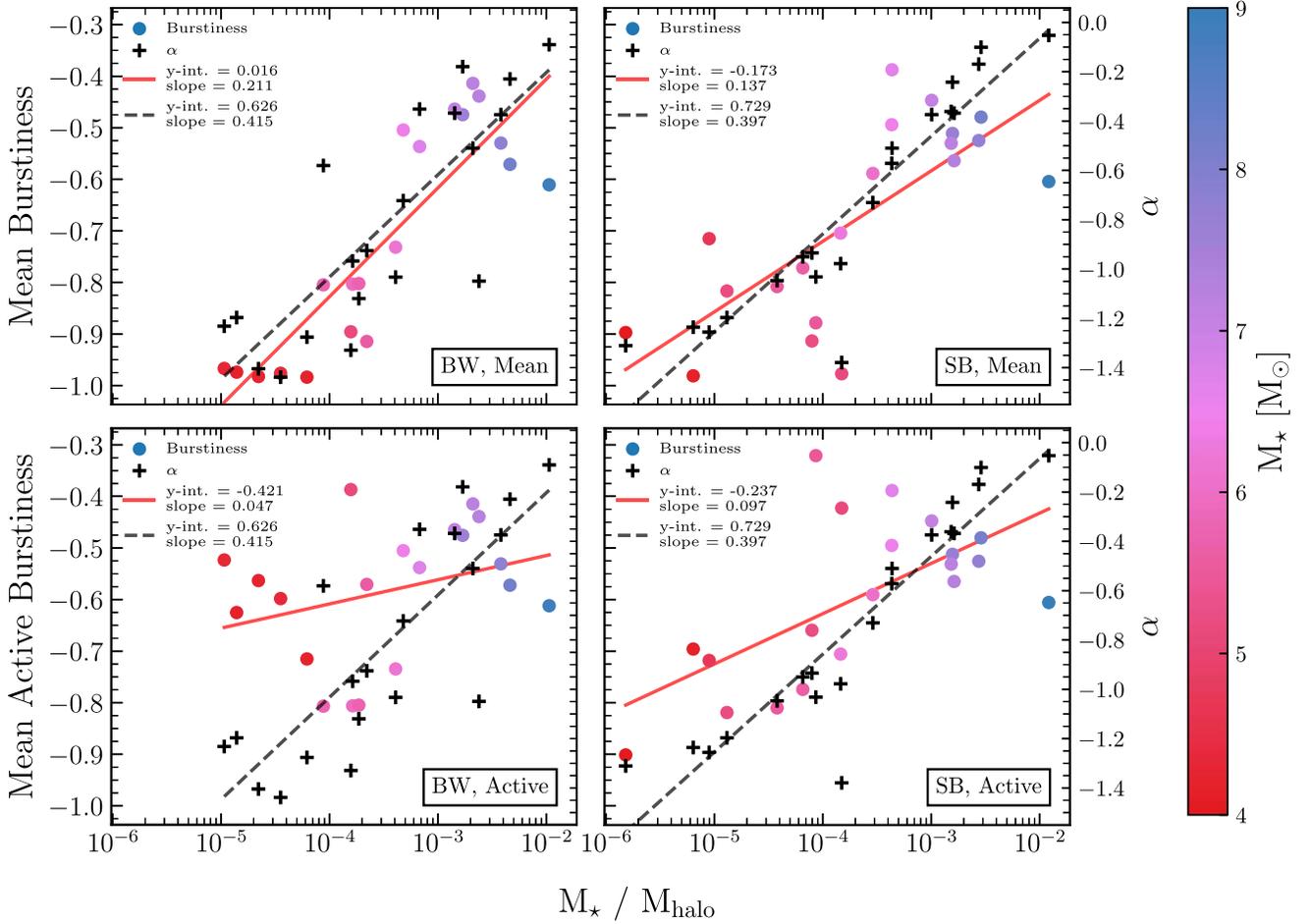

**Figure 7.** Burstiness and $\alpha[1\%-2\%]r_{vir}$ DM core slopes as a function of halo mass $M_*/M_{halo}$ for the Storm dwarf galaxies. Plots are colored by $M_* [M_\odot]$ where rows separate halos by simulation timescale (top: age of the universe; bottom: active star formation only) and columns separate halos by feedback model. Burstiness ranges from $-1$ to $1$ where a uniform distribution (SN rate = 0) has a burstiness $B = -1$ and an exponential distribution has a burstiness $B = 0$. The SN rate is calculated in 1 Myr increments for the age of the universe. Burstiness timing is averaged every 50 Myr beginning at every galaxy's star formation and ending at $z = 0$. Peak values of burstiness occur at lower values of $M_*/M_{halo}$ for the superbubble model, indicating that core formation may be less motivated by bursty star formation. For the active star formation timescale, the blastwave model shows higher burstiness values for the ultrafaint mass regime, indicating that UFDs are more sensitive to the subgrid physics of blastwave feedback.

influence the DM profiles by transforming from cuspier values to feedback-induced cores.

In Figure 7, the DM core slopes increase with the stellar mass fraction for both the superbubble and blastwave models. As cores are produced by SN energy driving repeated outflows, the relation between cores and stellar mass may indicate that feedback is underinjected or overinjected into the surrounding gas particles for the blastwave model. The superbubble feedback model shows a roughly linear relation between the stellar mass fraction and core slope, indicating that the subgrid physics produces core values more efficiently. To further illustrate the relation between core slopes and the stellar mass fraction, a linear regression model is used to calculate the y-intercepts, slopes, and the coefficients of determination ($R^2$) for burstiness and core slopes. The $R^2$ value gives the proportion of the variation in burstiness and core-slope values as predicted from the stellar mass fraction. $R^2$ ranges from 0 to 1, with 1 meaning that the statistical model has produced a good fit, or predicts the outcome in the linear regression setting.

Figure 7 shows that the average burstiness for both feedback models show a correlation with $\alpha$ core slopes as a function of the stellar mass fraction. $R^2$ is equal to 0.75 for burstiness in the blastwave feedback model, which indicates that burstiness depends strongly on the stellar mass fraction. $R^2$ is equal to 0.67 for the core slopes in the blastwave feedback model, indicating that the burstiness values are more motivated by the stellar mass fraction for this feedback model. For the superbubble model, $R^2$ is equal to 0.55 for the burstiness values, which indicates that burstiness has a lesser dependence on stellar mass. With $R^2$ equal to 0.85 for superbubble feedback, this suggests that the core slopes have a stronger correlation to $M_*/M_{halo}$ and are less motivated by bursty star formation.

The average burstiness during active star formation shown in Figure 7 shows a different outcome with the superbubble model resulting in a higher $R^2$ value for burstiness at 0.40 and 0.10 for blastwave feedback. The superbubble model shows less scatter between the core slopes and the stellar mass fraction during active star formation, further supporting that superbubble feedback may have a stronger dependence on $M_*/M_{halo}$.

Figure 8 shows the cumulative SFHs for each galaxy colored by $M_*$ in our Storm sample. Galaxies in the top row were simulated with blastwave feedback while galaxies in the bottom row implement superbubble feedback. Columns are ordered according to stellar mass, decreasing from left to right. Galaxies in the ultrafaint mass range form the majority of their stars in one or very few bursts of concentrated star formation spanning timescales shorter than 5 Gyr. Bright dwarfs and some classical dwarfs with higher stellar mass are star forming over the entirety of the simulation. The SFHs of the Storm halos indicate that stellar mass grows at a more uniform rate in more massive galaxies. As seen in Figure 7 the bluer points with elevated burstiness correspond to continuous SFHs. In contrast, galaxies which formed the majority of their stellar mass in 1–2 bursts of star formation have extremely low burstiness values. Star formation in two of the lowest luminosity superbubble dwarfs is not ignited until late times with the first stars forming at approximately 9 and 11.5 Gyr. In both cases the ignition of star formation is preceded by a significant interaction with a galaxy of a similar or larger size. One of the low-mass late-forming galaxies first started forming stars as it entered the virial radius of a larger host halo, indicating that ram pressure may have caused a sudden burst of star formation. Zhu et al. (2024) find that the SFRs of low-mass galaxies can be temporarily enhanced at infall due to mass flows driven by moderate levels of ram pressure. Additionally, in observations of large galaxy clusters Ebeling et al. (2014), Jáchym et al. (2019), and Poggianti et al. (2019) find that star formation can be triggered by ram pressure stripping. The second late-forming dwarf experienced a major merger with a gas-rich halo of a similar size over several gigayears leading up to its first stars forming. Zhang et al. (2020b, 2020a) and Gao et al. (2022) find evidence of efficient or enhanced star formation in the observed remnants of dwarf galaxy mergers, indicating that the merger event in our late star-forming dwarf could have induced star formation.

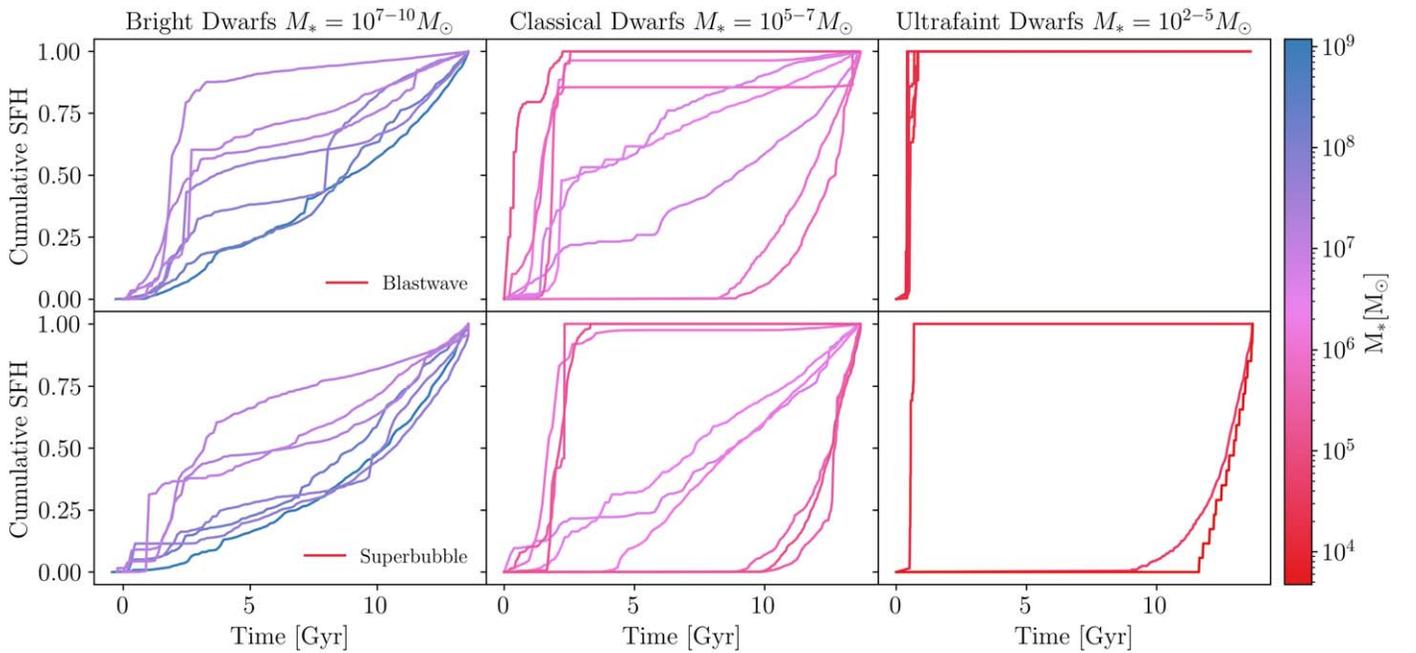

**Figure 8.** The cumulative SFH of each galaxy in our sample. The top row corresponds to halos run using blastwave feedback while the bottom row shows halos run using superbubble feedback. Plots are colored by $M_*$ and columns separate halos by stellar mass. The leftmost column contains the most massive dwarfs ($10^7$–$10^{10}$ $M_*[M_\odot]$), the middle column contains dwarfs in the stellar mass range $10^5$–$10^7$ $M_*[M_\odot]$, and the least massive dwarfs ($10^2$–$10^5$ $M_*[M_\odot]$) are shown in the rightmost column. These stellar mass ranges are classified as bright, classical, and ultrafaint, respectively, in accordance with Lazar et al. (2020).

## 4. Discussion

### 4.1. Metallicities

Calculating the average stellar and gas-phase metallicities of observed galaxies can be a powerful tool for gaining insight into the SFHs of galaxies in the Local Volume. As a result, comparing observed metallicity relationships to our sample is a great way to test how well the blastwave and superbubble feedback models regulate star formation. Feedback models that accurately reproduce observed metallicity relationships are assumed to roughly follow similar SFHs. Figure 2 shows that both the superbubble and blastwave samples are generally well constrained by the observations, indicating that both feedback methods effectively regulate star formation. In the ultrafaint mass range both feedback models diverge from the observed stellar metallicities. At this scale galaxies are too small to self-regulate, making them more susceptible to subgrid physics models (Munshi et al. 2019).

As seen on the left side of Figure 2 the stellar metallicities of galaxies from the MARVEL dwarf simulations have, on average, lower metallicity than the galaxies from the blastwave run of the Seven dwarfs simulations. There are two critical differences in the prescriptions for the stellar feedback and star formation models that could contribute to this decreased metallicity. The MARVEL dwarfs implement probabilistic star formation that depends on the $H_2$ abundance, density, and temperature of local gas particles (Christensen et al. 2012) while star formation in the Seven dwarfs solely depends on the temperature and density parameters. $H_2$ shielding produces large quantities of cold, clumpy gas that makes SN feedback more highly concentrated, improving feedback efficiency. Christensen et al. (2014) find that increased feedback results in higher SN mass loading factors and greater quantities of gas ejected from the central disk. Gas outflows remove metals from the disk, lowering stellar metallicity. Another key difference between the simulations is the quantity of energy associated with each individual SN is higher in the MARVEL dwarf simulation with blastwave feedback relative to the Seven dwarfs ($1.5 \times 10^{51}$ erg and $10^{51}$ erg, respectively). Galaxies from the Engineering Dwarfs at Galaxy Formation's Edge simulations with boosted SN energy were found to contain lower metallicity without substantially decreasing stellar populations (Agertz et al. 2020).

The average stellar metallicities of Storm galaxies are well constrained by observations in the most luminous galaxies, but at lower masses our sample diverges from the observed stellar metallicity relations, resulting in metal-poor, low-luminosity dwarfs. Many other hydrodynamical simulations systemically predict too low stellar metallicity measurements at a given luminosity (Simpson et al. 2013; Escala et al. 2018; Wheeler et al. 2019; Agertz et al. 2020; Applebaum et al. 2021), marking the effective retention of metals in the ISM as a major challenge in modern day simulation work. Currently IMF yields (Revaz & Jablonka 2018; Wheeler et al. 2019), insufficient time resolution (Macciò et al. 2017), preenrichment from a more massive host galaxy (Wheeler et al. 2019), and too efficient feedback (Agertz et al. 2020) are suggested as potential causes of lowered stellar metallicity in the faintest dwarfs.

There are a few proposed solutions to the problem of metal-deficient UFD populations such as implementing varying IMF yields or including metal-free (Population III) stars (Macciò et al. 2017; Revaz & Jablonka 2018; Wheeler et al. 2019; Prgomet et al. 2022; Sanati et al. 2023). Sanati et al. (2023) show that the inclusion of metal-free (Population III) stars with masses below 140 $M_\odot$ increases the average stellar metallicity in dwarf galaxies, though they find that their simulations still predicted galaxies with lower metallicities than those observed in the Local Group. Prgomet et al. (2022) find that incorporating a metallicity-dependent IMF results in top-heavy IMFs for low-metallicity ultrafaints. This boosts feedback and metal production in the faintest galaxies, lowering the stellar mass and increasing the average stellar metallicity.

Alternatively, Brown et al. (2019) find that the specific rates of Type Ia SNe and stellar mass are anticorrelated, indicating that Type Ia SN are central to the production of iron in UFDs. Gandhi et al. (2022) show that implementing a metallicity-dependent model for Type Ia SN rates better predicts the observed stellar mass–stellar metallicity relation. Time resolution limitations pose an additional challenge for effectively modeling Type Ia SNe in UFDs. Applebaum et al. (2021) suggest that it may be necessary to model "prompt" Type Ia SNe on timescales $\sim$100 Myr following the work of Mannucci et al. (2006) and Maoz et al. (2012) because the duration of star formation in UFDs can be much shorter than the $\sim$1 Gyr (Raiteri et al. 1996) currently implemented in the DCJL simulations.

### 4.2. Sizes

As shown by Figure 3 both the superbubble and blastwave feedback models produce galaxies that match the observed sizes of galaxies in the Local Group. There is no visual distinction between the galaxies from the Storm and Seven Dwarfs simulations, however, the halos from Wheeler et al. (2019) have much larger half-light radii at a given stellar mass than our sample.

According to Ludlow et al. (2019) simulations with star-to-DM particle mass ratios exceeding unity can be subject to spurious energy equipartition effects. Affected galaxies experience secular growth of sizes, increased velocity dispersions, and thicker disks resulting from gravitational scattering (Ludlow et al. 2020; Wilkinson et al. 2023). However, Ludlow et al. (2023) note that spurious heating is dominated by the amount of gaseous dissipation (i.e., the initial separation of DM and stellar particles in phase space). This implies that even in systems where the ratio of DM-to-stellar particle mass sizes exceeds unity, galaxies may not experience severe effects of energy equipartition unless the simulations also have a high amount of dissipation.

Energy equipartition effects are largely dependent on the choice of the softening length. Ludlow et al. (2020) find that reducing the gravitational softening length with fixed DM and gas particle masses can exacerbate the effects of two-body scattering, leading to systematically larger galaxy sizes. The DM particle masses in the highest-resolution simulations from Wheeler et al. (2019) are $\Omega_m/\Omega_b \sim 6$ times more massive than the masses of their star particles. These simulations also implement an adaptive force-softening length in their star-forming gas where the local density informs the choice of the softening length with a value of 3 pc for median ISM densities, decreasing to 0.1 pc in regions with high ISM densities where star formation typically takes place. Ludlow et al. (2020) found that for larger softening lengths galaxy sizes converged even when the star-to-DM particle mass ratios exceeded unity. It is likely that the larger softening length used in the MARVEL dwarf simulations minimizes mass segregation by energy equipartition despite the particle mass ratios exceeding unity.

### 4.3. Dark Matter Core Formation

We show that core formation is robust to the feedback model choice, with both the superbubble and blastwave simulations producing comparable cores. Additionally, we find that for values of $M_*/M_{\rm halo}$ between $2 \times 10^{-3}$ and $2 \times 10^{-2}$ both blastwave and superbubble feedback produce cored galaxies with peak core formation occurring at $M_*/M_{\rm halo} \sim 5 \times 10^{-3}$ in agreement with the fits produced by Di Cintio et al. (2014), Tollet et al. (2016), and Lazar et al. (2020). In contrast, Jackson et al. (2023) do not find a relationship between increasing core formation with $M_*/M_{halo}$ in the NewHorizon (NH) galaxies. They show that NH galaxies with extended, bursty SFHs exhibit cored profiles above a specific halo mass threshold ($\sim 10^{10.2}$ $M_\odot$) but also find that galaxies with cuspy central densities exist across a wide range of $M_*/M_{\rm halo}$ values. We find that the Storm simulations produce peak core values at lower $M_*/M_{\rm halo}$ ($\sim 5 \times 10^{-3}$) than the mass range and median $M_*/M_{\rm halo}$ value presented in Jackson et al. (2023). Additionally, Jackson et al. (2023) find that sustained star formation at the center of galaxies can increase the central potential and cause DM to cool adiabatically, reforming a cusp.

Jackson et al. (2023) indicate that the relation between $\alpha$ and $M_*/M_{\rm halo}$ may be obscured because NH galaxies overproduce stars; however, because core production occurs at similar halo and stellar masses relative to other simulations they argue that core formation is not solely dependent on the stellar mass normalized by the halo mass. We note that the core-slope values of the Storm simulations follow a trend of increasing core slope with $M_*/M_{\rm halo}$ that is consistent with other baryonic feedback simulations, and that there is less scatter in the relation.

Figure 6 illustrates the relationship between core slope and mean burstiness, showing less scatter for the superbubble sample than for the blastwave sample with standard deviation values of 0.19 and 0.21, respectively.

In the ultrafaint mass regime we find lower burstiness values, indicating that both models predict the lowest-mass dwarfs form the majority of their stellar populations in one large burst at high redshift, and are quenched in more recent years.

Blastwave burstiness reaches a maximum value at $M_*/M_{halo} \simeq 2 \times 10^{-3}$, just below the threshold of core formation. In the superbubble model, mean burstiness reaches a maximum value at a lower value of $M_*/M_{halo} \simeq 4 \times 10^{-4}$. These results indicate that core formation may be better predicted by the stellar mass fraction than burstiness for superbubble feedback. We note that the tight relationship between stellar and halo mass with core slope indicates that core formation may have a greater dependence on $M_*/M_{halo}$. We find that the superbubble model gives a less scattered relation overall between core slopes and $M_*/M_{halo}$ relative to the blastwave model, again indicating that core formation with superbubble feedback may have a stronger dependence on the stellar mass fraction.

We note that the superbubble model shows more scatter in mean burstiness for the UFD mass range over the entire timescale, which may be a consequence of feedback implementation. Superbubble feedback run with CHANGA is shown to reduce stellar mass in Milky Way–size galaxies relative to the blastwave model (Keller et al. 2015). Keller et al. (2014) and Mina et al. (2021) find that the SMHM relation for dwarf galaxies varies with superbubble feedback. A larger sample size may be necessary to determine systematic effects of superbubble feedback following the work of Munshi et al. (2019), which used 100 runs of an individual halo to study the effects of blastwave feedback on dwarf galaxy masses, and considers additional feedback details such as ionizing radiation, ISM turbulence, and runaway massive stars.

As seen in Figure 7, burstiness generally scales with the stellar mass fraction. For both SN feedback models UFDs form the bulk of their stars in very few major bursts at early times before becoming quenched. Bright dwarfs are seen to be star forming for much longer and are shown to have a more uniform distribution of star formation. Similarly, Emami et al. (2019) find that galaxies below $M_* \sim 10^{7.5} \, M_\odot$ undergo large, rapid bursts with timescales $\tau < 30$ Myr while galaxies above $M_* \sim 10^{8.5} \, M_\odot$ have smaller, slower bursts with timescales $\tau \gtrsim 300$ Myr.

As seen in Figure 8 simulated galaxies with elevated mean burstiness have continuous SFHs while the least bursty galaxies formed the majority of their stellar mass in 1–2 periods of star formation. Far-ultraviolet and H$\alpha$ emissions can be used to quantify the burstiness of observed galaxies (Sparre et al. 2017). Recently, Flores Velázquez et al. (2021) and Parul et al. (2023) used star formation indicators to measure the burstiness of galaxies from the FIRE project simulations. They found that Milky Way–mass galaxies have highly bursty star formation at early times and time-steady SFR at later times. Future work measuring the SFHs of galaxies in the local Universe could constrain the burstiness of dwarf galaxies and provide a useful comparison to the results of this paper.

Improving the observed stellar mass fraction as a function of halo mass will also help to constrain early stellar feedback (Read & Erkal 2019; Revaz & Jablonka 2018). Smith (2021) finds that adopting IMF-averaged rates produces simulations that overestimate the impact of radiation from photoionization. However, when compared with explicitly sampling the IMF, an averaged sampling was found to trigger SNe identically in the case where SNe were the only source of feedback.

Additionally, Applebaum et al. (2020) and Gurvich et al. (2023) show that a stochastic IMF results in burstier feedback that quenches star formation at earlier times for UFDs. As discussed in Munshi et al. (2019, 2021), many of the low-mass halos in our H$_2$ star formation model are not able to form stars before reionization prevents them from doing so. This implies that our results are sensitive to our selected reionization model. We have adopted the UV background model of Haardt & Madau (2012), which has been shown to heat the IGM earlier ($z \sim 15$) than it should (Oñorbe et al. 2015), potentially making the impact of reionization particularly strong on our results.

A more realistic reionization model may allow dark halos in the lower-density regions of our models to form stars, allowing UFD halos to experience longer quenching times. Benitez-Llambay & Frenk (2020) explicitly look at the effect of reionization on the $z = 0$ occupation fraction, and find that it varies with how early or late reionization begins. Understanding the impact of reionization will require further simulations and study, but will be essential to explore the uncertainties that will impact the interpretation of VRO's LSST observations. We note that both versions of the Storm simulation were run with the identical UV background, making the comparison of the SN feedback model robust.

Ultimately, we find the Storm simulations produce consistent results for higher dwarf galaxy masses regardless of the feedback model. The relation between burstiness and the DM core slopes remains robust for both models and we detect similar efficiencies for the superbubble model within this mass range during active star formation. We also find that UFDs exhibit similar modeling difficulties that are present in the existing literature and will need to be constrained by future observations by VRO and the Nancy Grace Roman Space Telescope.

## 5. Conclusion

We compare the observed properties of galaxies produced using two distinct SN feedback models (blastwave and superbubble) down to the ultrafaint mass range. We have adopted the burstiness parameter defined by Applebaum et al. (2020) to compare subgrid physics and the relationship between burstiness and central DM density slopes for the Storm simulation using both the superbubble and blastwave feedback models. Implementation of baryonic processes with the high resolution of the MARVEL-ous Storm simulations allows for improved analysis of a full range of dwarf galaxies from LMC mass down to the ultrafaint mass regime. The simulation is able to produce feedback-induced cores that strongly depend on the ratio of stellar mass to halo mass regardless of feedback model.

1. We confirm that the blastwave feedback model generally produces elevated stellar masses and higher burstiness values, as shown in Figure 7, and that the superbubble model shows evidence for stronger feedback despite having lower values of energy per SN event. We find that this is consistent with the recent work by Mina et al. (2021), which shows superbubble feedback produces stronger outflows that expel high quantities of low angular momentum gas and regulate star formation.

2. We detect no visible distinction in the the gas phase and average stellar metallicities between the feedback models for the Storm galaxies in Figure 2. We also note that the Storm galaxies have lower stellar metallicities than those produced with blastwave feedback in other high-resolution ΛCDM simulations (Mina et al. 2021) due to the elevated SN feedback energy and the $H_2$ star formation model used in our simulation.

3. The superbubble and blastwave feedback models in the Storm simulations were found to produce galaxies that are consistent with the sizes of observed galaxies as seen in Figure 3. Our results are in disagreement with Mina et al. (2021), who found that the superbubble model produces galaxies with increased half-light radii. The UFD galaxies from Wheeler et al. (2019) are much larger in size, which may be related to the energy equipartition effects described in our Discussion.

4. We confirm that continuous bursty star formation is crucial for sustained core formation. Figure 7 shows the relation between burstiness and core slope is robust for both the blastwave and superbubble feedback models in the higher dwarf galaxy mass regime. UFDs have cuspier density profiles indicating a lack of sustained feedback to produce cores, as expected from the limitation on reionization and subsequent UFD modeling within our simulations.

5. We find that effective core formation peaks at $M_*/M_{\rm halo} \simeq 5 \times 10^{-3}$ for both feedback models with mean burstiness generally tracking core formation in Figure 7. Maximum burstiness values occur at $M_*/M_{\rm halo} \simeq 2 \times 10^{-3}$ for the blastwave feedback and $M_*/M_{\rm halo} \simeq 4 \times 10^{-4}$ for superbubble feedback. This lower value of $M_*/M_{\rm halo}$ for superbubble feedback indicates that core formation may be less motivated by burstiness, and that the stellar mass fraction may better predict whether a dwarf galaxy is able to sustain a core. Superbubble feedback produces a more scattered relation between $M_*/M_{\rm halo}$ and average burstiness in the UFD regime, as shown in Figure 7.

Future work will investigate the burstiness–core slope relation with respect to $M_*/M_{\rm halo}$ by examining the effects of the star formation criteria and star formation efficiency on various properties of the MARVEL galaxies to determine any interactions with feedback that may result in a stronger correlation with core-slope formation. We also acknowledge that the results presented in this paper are only dependent on the stellar feedback prescriptions and may be altered with the addition of other feedback channels. SN events emit radiation via UV ionizing photons that influence the environment in which star formation and SN detonation occurs. Hopkins et al. (2020) show that radiative feedback influencing the UV background is crucial to suppressing star formation in dwarf galaxies, and can affect the burstiness in galaxies without the implementation of radiation feedback to smooth SFHs. Future work will incorporate this radiation feedback to understand the interplay between radiation and traditional stellar feedback models, as well as determine any effects on the burstiness, SFHs, and DM properties of dwarf galaxies. With their addition, the treatment of radiative feedback becomes crucial when considering that the ISMs in simulated galaxies are greatly influenced by feedback processes (Agertz & Kravtsov 2015). Star formation in galaxy simulations cycles through the development and termination of cold gas as it interacts with stellar feedback. Star formation thus becomes regulated as it cycles through weakened feedback and dense gas to form more stars. To prevent the overproduction of stars, the elimination of cold gas requires simulations to unrealistically model the amount of gas in the warm phase (Agertz et al. 2020). To understand the relation of strong stellar feedback and self-regulation, it becomes necessary to determine the effects of radiation feedback on dwarf galaxy properties and whether or not we can produce a realistic ISM. In addition, we will explore instantaneous burstiness within the context of reionization, as lower-mass halos are unable to form stars due to the nonequilibrium $H_2$ star formation prescription used in the MARVEL simulations. Additional work is in progress to determine any distinguishable effects of the blastwave and superbubble feedback models on the CGM.

## Acknowledgments

A.C.E. is supported by the National Aeronautics and Space Administration Future Investigators in NASA Earth and Space Science and Technology (FINESST; 22-ASTRO22-0096). J.D.V. is supported by the Bullard Dissertation Completion Fellowship from the University of Oklahoma. Long before the University of Oklahoma was established, the land on which the University now resides was the traditional home of the Hasinais Caddo Nation and Kirikiris Wichita and Affiliated Tribes; more information can be found at https://www.ou.edu/cas/nas/land-acknowledgement-statement. F.D.M. acknowledges support from NSF grant PHY2013909. F.D.M. is grateful for the hospitality of Perimeter Institute where part of this work was carried out. Research at Perimeter Institute is supported in part by the Government of Canada through the Department of Innovation, Science and Economic Development, and by the Province of Ontario through the Ministry of Colleges and Universities. This research was also supported in part by the Simons Foundation through the Simons Foundation Emmy Noether Fellows Program at Perimeter Institute. C.R.C. was supported by the NSF CAREER grant AST-1848107.

## ORCID iDs

Bianca Azartash-Namin https://orcid.org/0000-0002-7898-6194
Anna Engelhardt https://orcid.org/0009-0006-2183-9560
Ferah Munshi https://orcid.org/0000-0002-9581-0297
B. W. Keller https://orcid.org/0000-0002-9642-7193
Alyson M. Brooks https://orcid.org/0000-0002-0372-3736
Jordan Van Nest https://orcid.org/0000-0003-3789-3722
Charlotte R. Christensen https://orcid.org/0000-0001-6779-3429
Tom Quinn https://orcid.org/0000-0001-5510-2803
James Wadsley https://orcid.org/0000-0001-8745-0263